\documentclass{kapproc} % Computer Modern font calls

\usepackage{procps} 

\usepackage[dvips]{graphicx}

\upperandlowercase
\kluwerbib % will produce this kind of bibliography entry:

\begin{document}

\articletitle{X-rays from disk galaxy halos, Ly$\alpha$
  from forming galaxies,
  and the $z\sim$~1 TF relation}

%\articlesubtitle{This is an Article Subtitle}

\author{Jesper Sommer-Larsen}
\affil{Dark Cosmology Centre, Niels Bohr Institute,\\ 
University of Copenhagen, Juliane Maries Vej 30, DK-2100 Copenhagen, Denmark}
\email{jslarsen@tac.dk}

\begin{abstract}
Extended, soft X-ray emission from the halo of a very large disk
galaxy has been detected. The luminosity and surface brightness
distribution is in excellent agreement with predictions by recent,
cosmological galaxy formation models. Predicted Ly$\alpha$ emission,
associated with ``cold'' accretion of filamentary gas onto galaxies,
is discussed in relation to Ly$\alpha$ ``blobs''. Finally, the
predicted evolution of the Tully-Fisher relation, going from $z$=0 to
1, is discussed in relation to recent observations.

\end{abstract}


\section{X-ray emission from disk galaxy haloes}
\begin{figure}[ht]
\hfill\includegraphics[angle=0,width=2.3in]{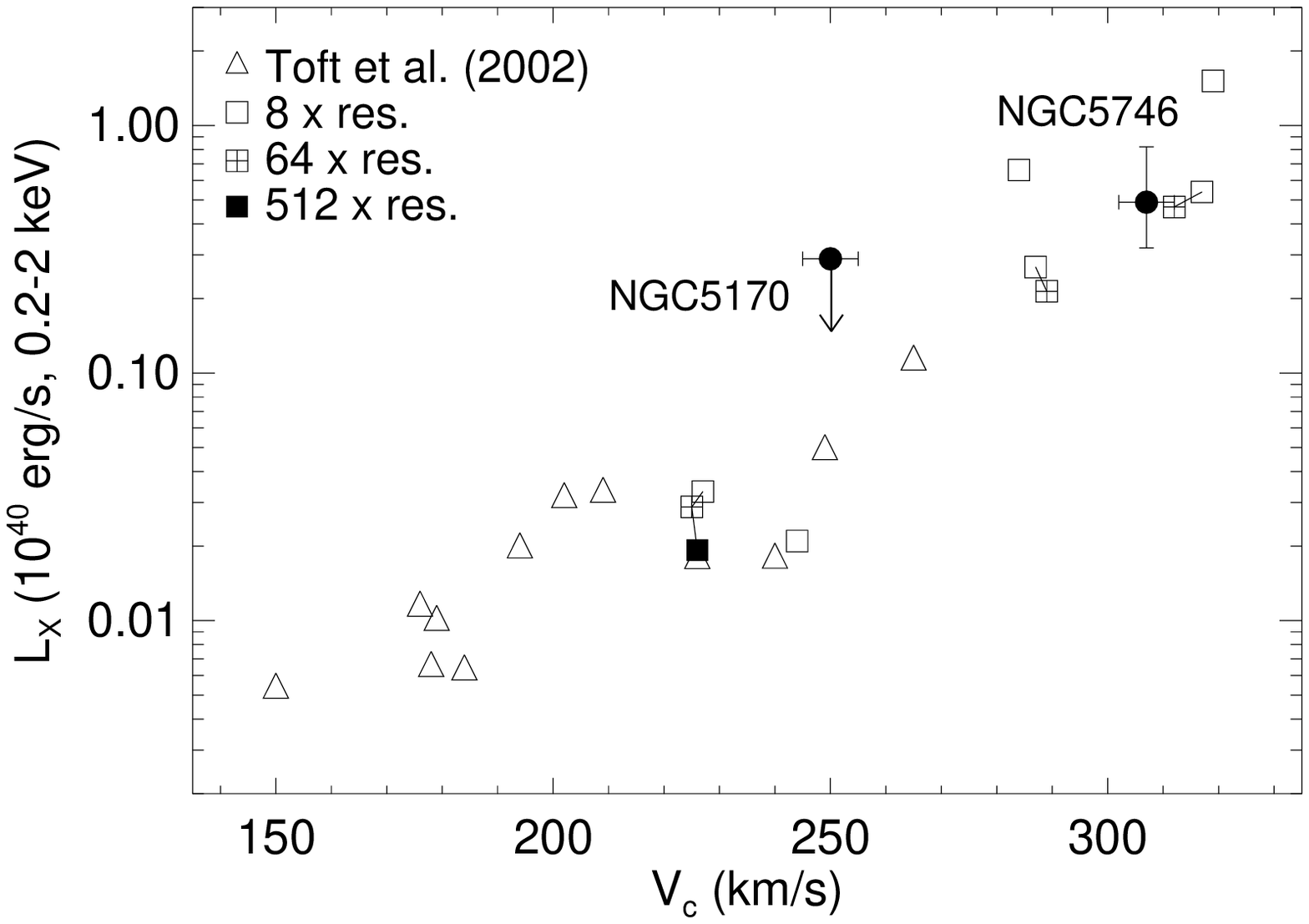}
\hfill\includegraphics[angle=0,width=2.3in]{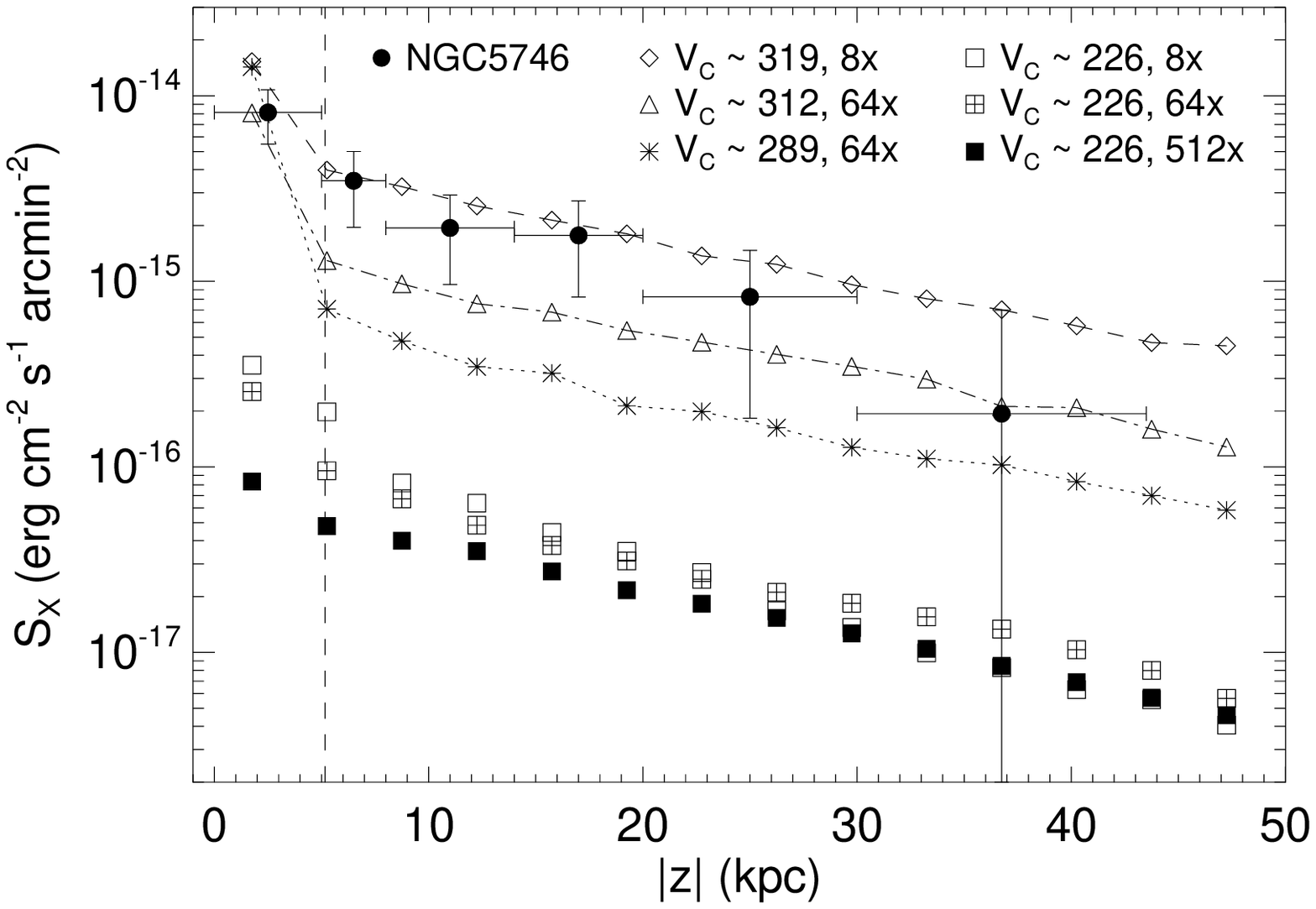}\hspace*{\fill}
\caption{Left: Predicted and observed 0.2-2 keV luminosities of
X-ray haloes as a function of disc circular velocity. All X-ray
luminosities have been calculated within the same physical aperture as
used for NGC 5746. The filled circles are from the observations of NGC
5746 and NGC 5170 (1-$\sigma$ upper limit of $L_X<2.9~10^{39}$ erg/s,
0.3-2 keV), while other
symbols are the predictions from simulations with a range of different
resolutions and circular velocities: Triangles are for simulations
with low-metallicity chemical composition while squares are for
simulations with self-consistent chemical evolution run at 8, 64,
and 512 times the original resolution (Sommer-Larsen, Gotz and
Portinari 2003) corresponding to gas
particle masses of 7.5~10$^5$, 9.4~10$^4$ and 1.2~10$^4~h^{-1}M_{\odot}$,
respectively ($h$=0.65). Results from simulations re-run at
higher resolution are connected with lines. Open squares, except the
simulated galaxy with a circular velocity of $\sim$225 km/s, are for
simulations run with a universal baryon fraction of 0.15. All other
simulations were run with a baryon fraction of 0.1. Right:  Predicted
and observed surface brightness profile of X-ray haloes as function of
the distance to the disc mid-plane. Filled circles are NGC 5746 data
while other symbols mark simulations with different resolutions and
circular velocities. The vertical dashed line indicates D25 of NGC
5746.}
\end{figure}
Disk galaxies continue forming to the present day, as
evidenced by, e.g., infall of high-velocity clouds and satellite
galaxies. Self-consistent models of disk galaxy formation
predict that, at present, part of the inflowing gas originates as hot
and dilute, low-metallicity halo gas, slowly cooling out and accreting
onto the disk (e.g., Abadi et al. 2003, Sommer-Larsen et al. 2003,
Governato et al. 2004, Robertson et al. 2004). The X-ray luminosity of
the hot halo is predicted to increase strongly with galaxy mass, and the
haloes of the most massive galaxies should be detectable 
(Toft et al. 2002, Rasmussen et al. 2004). 
Searches for this hot halo gas have so far
failed to detect such soft X-rays, challenging galaxy formation theory. 
Moreover, it has been suggested
that for galaxies of total mass less than a few times 10$^{11}
M_{\odot}$,
most gas stays cold during the accretion onto the galaxy (apart from
a very transient radiative shock phase, e.g., Birnboim \& Dekel 2003; 
Dekel \& Birnboim 2005; next section). This may result in reduced
halo X-ray emission.
On the other hand, the recent detection of a
warm-hot phase of the intergalactic medium shows the presence of a
reservoir of hot and dilute gas at galactic distances (Nicastro et
al. 2005). Furthermore,
absorption of the OVI line in quasar spectra (Wakker et al. 2004), and the
head-tail structure of some high-velocity clouds in the
halo of the Milky Way (Bruns et al. 2000) provide circumstantial
evidence  that the Milky Way is surrounded by an extended hot halo.

As a test of current disk galaxy formation models we used the Chandra X-ray
telescope to study the most promising candidate spiral
galaxy for detecting halo X-ray emission, NGC 5746. We also studied a
similar, but less massive galaxy, NGC 5170, as a test of our
procedure. The galaxies are massive and nearby (NGC 5746 is an SBb
galaxy at $d$=29.4 Mpc and has $V_c$=307$\pm$5
km/s; NGC 5170 is an Sc galaxy at $d$=24.0 Mpc and has $V_c$=250$\pm$5
km/s). Both galaxies are quiescent, showing no signs of either
starburst activity, interaction with other galaxies, or an active galactic
nucleus. The disks of the galaxies are viewed almost perfectly
edge-on.
 
Diffuse, soft X-ray emission extending more than 20 kpc from the
stellar disc was detected around NGC 5746. A total of
about 200 net counts in the 0.3-2 keV band were detected from the halo
of  NGC 5746, corresponding to a 4.0-$\sigma$ detection. The same
observing technique and data analysis revealed no diffuse emission
around the less massive galaxy NGC 5170. Moreover, from Fig.1 it
follows that the agreement between observations and models is very good.
Full discussions of the results are presented in Pedersen et
al. (2005), and Rasmussen et al. (2005).

\section{``Cold'' accretion, and Ly$\alpha$ properties of forming galaxies}
\begin{figure}[ht]
\hfill\includegraphics[angle=0,width=2.3in]{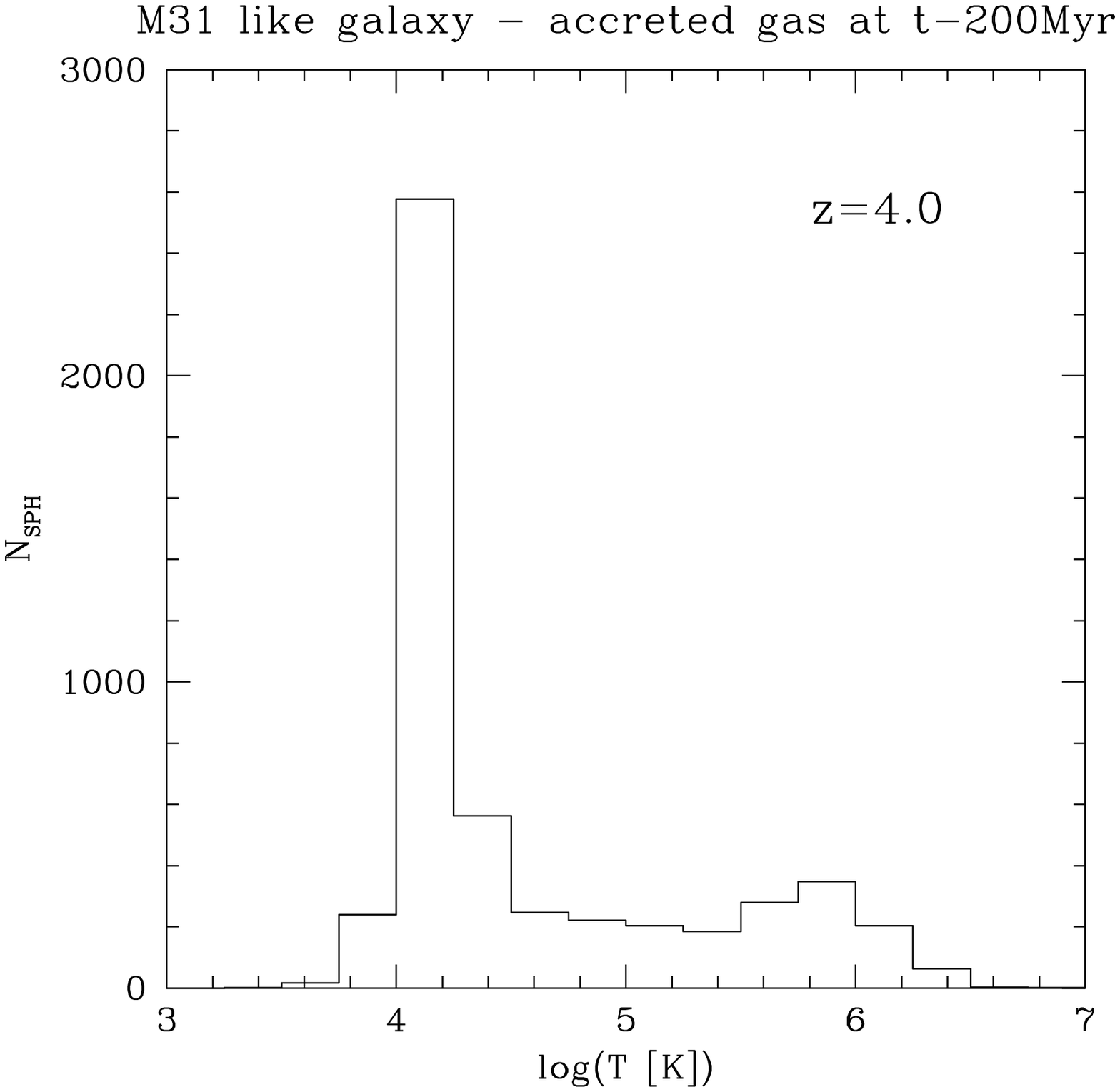}
\hfill\includegraphics[angle=0,width=2.3in]{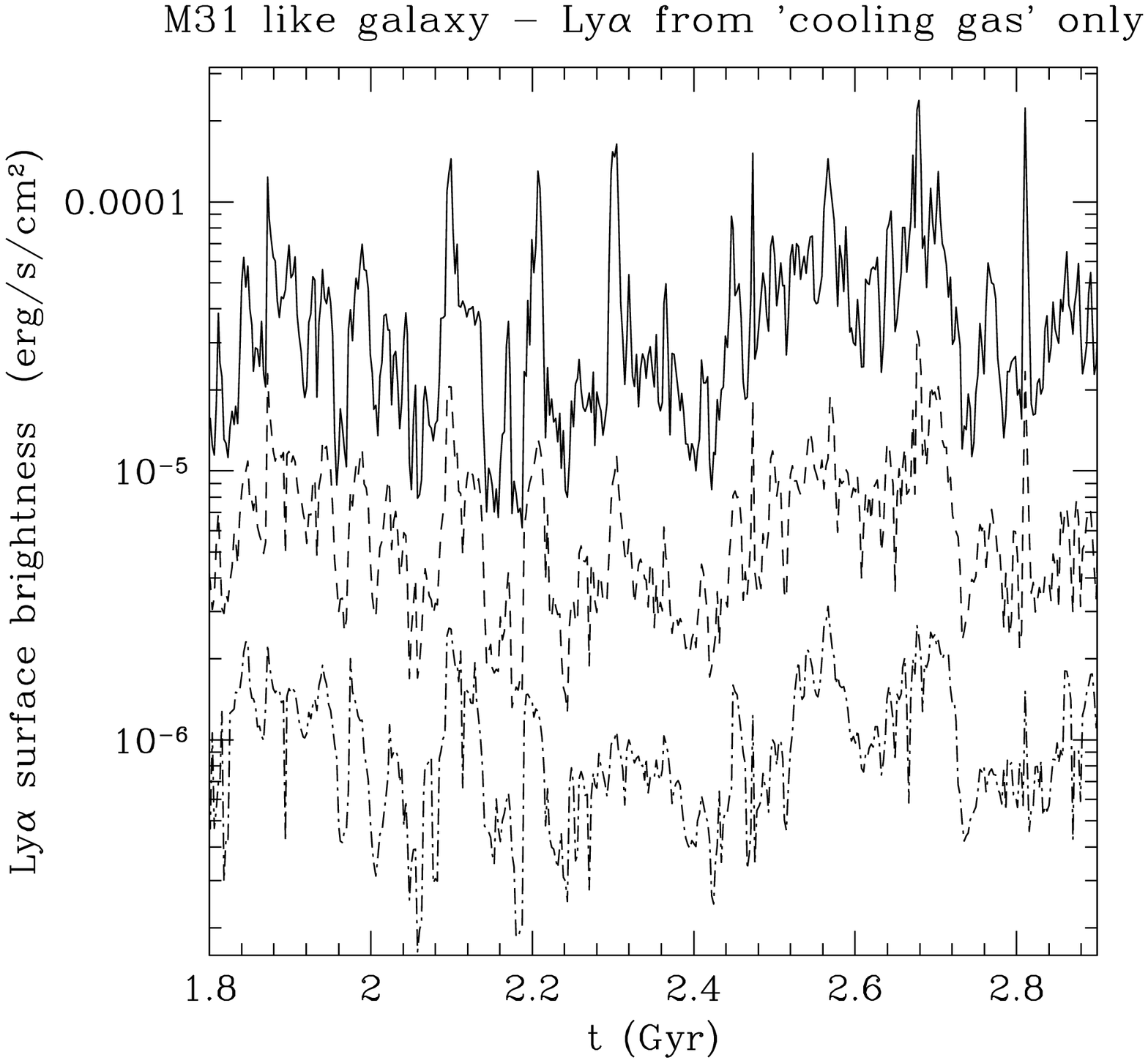}\hspace*{\fill}
\caption{Left: Temperature distribution of gas particles, prior to
  accretion onto a number of starforming proto-disks in a M31 like 
  proto-galaxy. Shown is (at a time 200 Myr before the $z$=4 frame)
  temperatures of gas particles, which
  at $z$=4 are located in the starforming proto-disks at densities
  $n_H>$ 10 cm$^{-3}$, and which 200 Myr prior to this have not been
  accreted onto the proto-disks yet. Right: Ly$\alpha$ surface
  brightness of ``cooling'' radiation from the region around a M31
  like proto-galaxy at $z\sim$3, versus time. Top curve shows the peak
  surface brightness; second and third averages over circular
  regions of radii 10 and 20 kpc, respectively, centered at the
  position of the peak Ly$\alpha$ surface brightness.}
\end{figure}
Not all gas, 
ending up as cold and star-forming in galaxies, has been shock-heated to 
temperatures $T\sim10^6$ K, and then (considerably) later accreted onto 
the galaxy. Rather, some of the gas is accreted at densities and rates, 
such that accretion shocks are strongly radiative, and the gas remains at 
$T\sim10^4$ K during the accretion, cooling mainly by
abundant Ly$\alpha$ emission (previous section). 
So, although most of the {Ly$\alpha$ photons produced in young 
galaxies originate
from photo-ionized HII regions around young, massive stars, 
a fraction of the Ly$\alpha$ photons
produced in and around the galaxies originate from cooling, almost 
neutral gas -- we find this fraction to be typically of order 10\%.
In Fig.2 (left) is shown, for a high-resolution galaxy formation
simulation (2.2 million particles, $m_{\rm{gas}}=9.4~10^4~h^{-1}M_{\odot}$), 
the temperature distribution of gas before it
is accreted onto the proto-disks of a M31 like proto-galaxy, at $z$=4. As
can be seen, at this redshift, the majority of the gas is, in fact,
accreted ``cold''.

Can Ly$\alpha$ emission associated with ``cold'' accretion be observed?
In recent years a number of so-called Ly$\alpha$ ``blobs'' have been
detected (Keel et al. 1999, Steidel et al. 2000, Francis et al. 2001, 
Matasuda et al. 2004, Palunas et al. 2004, Dey et al. 2005).
These are systems with spatial extents of up to 100 kpc, and
Ly$\alpha$ luminosities of up to 5~10$^{43}$ erg/s. In all cases, but
one (see below), counterparts (like optical, IR, X-ray or radio) have
been detected. Various mechanisms
have been proposed, like i) QSO illumination (e.g., Haiman \& Rees 2001;
Weidinger et al. 2004), ii) galactic super-winds (e.g., Mori et al. 2004;
Wilman et al. 2005), or iii) cold accretion. In the case of a $z$=3.15
blob, discovered by Nilsson et al. (2005), there are no other sources
associated with the object going from X-rays (Chandra) to 8~$\mu$m emission
(Spitzer). The blob has a linear extent of about 50 kpc and a typical
surface brightness of 4~10$^{-4}$ erg/s/cm$^2$, at the source. Moreover,
a galaxy of photometric redshift 3.0 is situated at a projected distance of
about 40 kpc from the blob, and may, given the uncertainty in the
photometric redshift, be physically associated with the blob.

To test whether this blob can be filamentary gas being accreted
``cold'' onto a companion galaxy, we conducted the following experiment:
for a M31 like proto-galaxy we calculated the Ly$\alpha$ surface brightness,
in a 100x100 kpc (projected) region centered on the proto-galaxy,
of ``cooling''
radiation only (so all contributions from regions with young stars
were  removed, as well as all emission, in general, from gas closer
than 10 kpc to any star-forming region). The result is shown in Fig.2 
(right), at $z\sim$3, for a period of about 1 Gyr, with time
resolution of just 2.5 Myr. As can be seen from the figure, we can get to
within about an order of magnitude of the observed surface brightness
level. This is interesting, and may point to a cold accretion origin
of the blob Ly$\alpha$ emission, just on a larger scale, such as
filamentary gas accretion onto a galaxy group -- this option is
currently being investigated. Given that in a search volume of about
40000 comoving Mpc$^3$, only one such blob has been detected, it is
actually comforting, that we could {\it not} reproduce the blob
characteristics, by cold accretion onto this, randomly selected, M31
like galaxy. 
    
\section{The $z\sim1$ Tully-Fisher relation}
\begin{figure}[ht]
\includegraphics[width=4.0in]{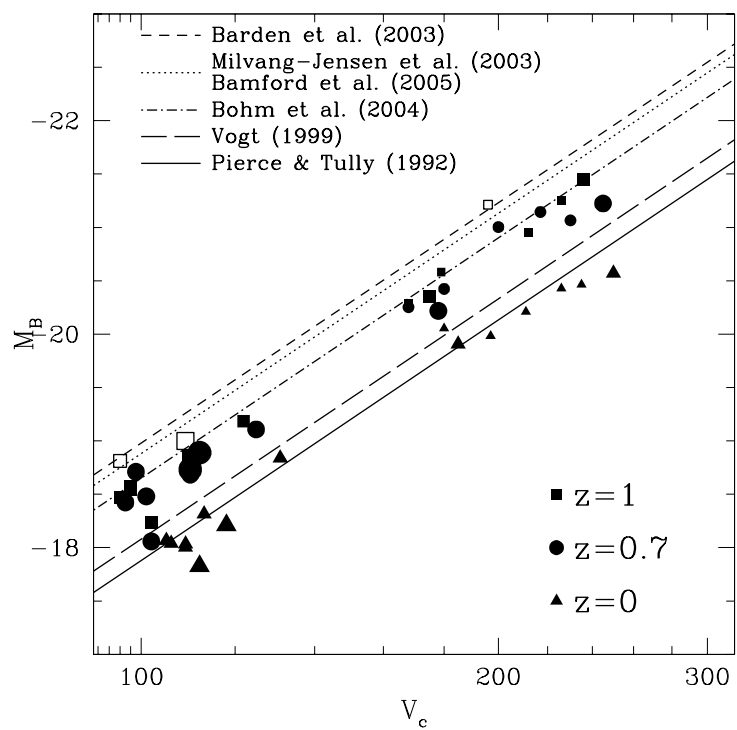}
\caption{Tully-Fisher relations from observations, as well as,
  simulations. Solid line: B band $z$=0 TF relation of Pierce \& Tully
  (1992) for Sc galaxies (B-V$\sim$0.55), shifted downwards by 0.6 mag
  to enable a comparison to the model disk galaxies which have
  B-V=0.6-0.75 ($\sim$Sb type). Other lines show the differential
  evolution of the zero-point of the B-band TF going to $z\sim$~1, 
  reported by various authors (the slopes have been forced to be as
  for  the $z$=0 TF relation). Results of models, invoking feedback,
  self-consistent treatment of chemical evolution and metal-dependent
  radiative cooling, and assuming a universal baryon fraction of 0.15
  (Portinari \& Sommer-Larsen 2005) are shown by
  various symbols. Filled symbols correspond to $z$=0, 0.7 and 1. Open
  symbols refer to
  disk galaxies, which were too disturbed at $z$=1 to enable
  determination of the TF quantities; hence these were determined at
  $z$=0.8 instead. Smallest symbols correspond to the original
  resolution of Sommer-Larsen et al. (2003), medium sized and large to
  8 and 64 times higher mass resolution, respectively.}
\end{figure}
Considerable effort has been spent in recent years to determine the
evolution of the zero-point and slope of the disk galaxy Tully-Fisher relation
going from $z$=0 to $z\sim$~1 (e.g., Vogt 1999; Barden et al. 2003,
Milvang-Jensen et al. 2003; Bohm et al. 2004; Bamford et al. 2005). 
The observational results 
span a range from almost no evolution to about 1.1 mag 
fading in the B band. Such observational information provides
important constraints on models of galaxy formation and evolution, in 
particular no or only very mild evolution points to scenarios with
considerable, continuing disk growth, going from $z$=1 to 0.

Given this, it is clearly of interest to
compare to what fully cosmological galaxy formation simulations,
incorporating feedback, non-instantaneous chemical evolution,
metal-dependent radiative cooling etc. predict. We have recently
engaged in such a study, and show the first results in Fig.3; the full
results will be presented in Portinari \& Sommer-Larsen 2005. The
preliminary conclusion based on the results shown in Fig.3 is that our
results for $z$=0.7 and 1.0, relative to the $z$=0 results are nicely
bracketed by the various observational determinations at $z\sim$~1.
Moreover, we find that the stellar {\it mass} TF relation shows
essentially no evolution going from $z$=1 to 0, since an increase in
stellar mass leads to an increase in $V_c$ as well. This is in
agreement with the findings of Conselice et al. (2005).

\begin{acknowledgments}
I thank my collaborators on the various projects for
providing material to this contribution.
\end{acknowledgments}

\end{document}